%% file: paper.tex
\documentclass[sigconf]{acmart}
\usepackage[utf8]{inputenc}

\usepackage{hyperref}
\usepackage{tabularx}
\usepackage{booktabs}
\usepackage{soul}
\usepackage{xcolor}
\usepackage{xspace}
\usepackage{graphicx}
\usepackage[caption=false]{subfig}  
\usepackage[export]{adjustbox}
\usepackage{amsmath}
\usepackage{textcomp}
\usepackage{listings}
\usepackage{algorithm}
\usepackage[noend]{algpseudocode}
\usepackage{balance}
\usepackage{enumitem}
\usepackage{multirow}
\usepackage{colortbl}
\usepackage{balance}
\usepackage[flushleft]{threeparttable}

\definecolor{codegreen}{rgb}{0,0.6,0}
\definecolor{codegray}{rgb}{0.5,0.5,0.5}
\definecolor{codepurple}{rgb}{0.58,0,0.82}
\definecolor{backcolour}{rgb}{0.95,0.95,0.92}
\definecolor{gray}{gray}{0.9}

\lstdefinestyle{mystyle}{
    stringstyle=\color{deepgreen},
    backgroundcolor=\color{backcolour},   
    commentstyle=\color{codegreen},
    keywordstyle=\color{magenta},
    numberstyle=\tiny\color{codegray},
    stringstyle=\color{codepurple},
    basicstyle=\ttfamily\footnotesize,
    breakatwhitespace=false,         
    breaklines=true,                 
    captionpos=b,                    
    keepspaces=true,                 
    numbers=left,                    
    numbersep=5pt,                  
    showspaces=false,                
    showstringspaces=false,
    showtabs=false,                  
    tabsize=2
}

\newcommand{\WA}{WAs\xspace}

\setcopyright{none}
\renewcommand\footnotetextcopyrightpermission[1]{}

\begin{document}

\title{What Writing Assistants Can Learn from Programming IDEs}

\author{Sergey Titov}
\affiliation{
    \institution{JetBrains Research}
    \city{Paphos}
    \country{Republic of Cyprus}
    }
\email{sergey.titov@jetbrains.com}

\author{Agnia Sergeyuk}
\affiliation{
    \institution{JetBrains Research}
    \city{Belgrade}
    \country{Republic of Serbia}
    }
\email{agnia.sergeyuk@jetbrains.com}

\author{Timofey Bryksin}
\affiliation{
  \institution{JetBrains Research}
    \city{Limassol}
    \country{Republic of Cyprus}
    }
\email{timofey.bryksin@jetbrains.com}

\begin{abstract}
    With the development of artificial intelligence, writing assistants (WAs) are changing the way people interact with text, creating lengthy outputs that can be overwhelming for users. The programming field has long addressed this issue, and Integrated Development Environments (IDEs) have been created for efficient software development, helping programmers reduce the cognitive load. This experience could be employed in the development of WAs. IDEs can also be used to test assumptions about interventions that help people interact with WAs efficiently. Previous works have successfully used self-written IDE plugins to test hypotheses in the field of human-computer interaction. The lessons learned can be applied to the building of WAs.
\end{abstract}

\maketitle
\pagestyle{plain}

\input{sections/01-intro.tex}
\input{sections/02-body.tex}
\input{sections/03-conclusion.tex}

\bibliographystyle{ACM-Reference-Format}
\balance
\bibliography{paper}

\end{document}

%% file: sections/01-intro.tex
\section{Introduction}\label{sec:intro}

Writing assistants (\WA) are changing the way people interact with text. These tools can be incredibly useful in improving the quality and efficiency of writing~\cite{gero2022design}. However, with recent progress in the development of large language models such as GPT-3~\cite{brown2020language} and MT-NLG~\cite{smith2022using}, \WA can create long output~\cite{shi2022effidit} that can sometimes be overwhelming and difficult to manage for anyone and especially for neurodiverse users. Creating tools that aim to reduce cognitive load during writing-related tasks can help address this problem. 

The programming field has long been working on this issue. Code writing is a mentally challenging task that imposes much cognitive load on the user. For this reason, Integrated Development Environments (IDEs), an ecosystem of \WA, have been created to make code writing more efficient and less demanding for mental effort~\cite{habeck2008security,fakhoury2020measuring,hunter2021ten}. Over the last several decades, a lot of work has already been done to support users' performance by incorporating programming language peculiarities into IDEs (\textit{e.g.}, keyword coloring, code writing conventions, and fixed context). However, there is still a long way to go. To embrace inclusivity, thorough research should be carried out on possible interventions to reduce the cognitive load of various groups of users. As an ever-changing industry-level tool, an IDE is an excellent instrument for that: tasks, changes, and measurements can be easily incorporated into the system by plugins to test hypotheses~\cite{denissov2021creating,ramos2022tool}.

We argue that established practices in the development of IDEs, as well as the use of IDEs to research \WA, can be fruitful for next-generation \WA development.

%% file: sections/02-body.tex
\section{IDE Experience For Writing assistants}\label{sec:arg}

\subsection{IDE's ways to reduce cognitive load}

IDEs are advantageous for efficient software development. To provide assistance, IDEs leverage the hallmarks of programming languages and development practices. The usefulness of this could be assessed, since IDEs allow us to collect various metrics that, in turn, reflect users' efficiency and other characteristics of their performance~\cite{biehl2007fastdash,bhardwaj2011redprint}. 

In this section, we discuss some techniques that are used in IDEs to help programmers reduce the cognitive load and suggest how these techniques could be used in the next generation of \WA in general writing domains.

\textbf{Code coloring.}
Unlike other texts, code has a structure-based readability improvement: the coloring of keywords and literals (\textit{e.g.}, strings or numbers). IDEs take advantage of this by using color coding with respect to the programming language's syntax. That makes it easier for programmers to quickly recognize the structure of the code snippet and follow its logic. Although structural word coloring is mostly unique to programming languages, there is some design space where we can use it in AI-based \WA. For example, various machine learning explainability tools use coloring to help the user understand and interpret the results of the model~\cite{ribeiro2016should,NIPS2017_7062}. Similarly, in \WA, colors can be used to highlight key parts of a paragraph that are most important to convey a specific message or idea. Such a feature might be especially useful in academic and law writing, where the coherence of argumentation is critical. In those domains, word coloring would make it easier for a writer to organize their thoughts.

\textbf{Code style conventions.}
Coding style is a formally defined set of conventions that helps to support consistency in a codebase and makes it easier to comprehend. As a result, these conventions can affect both the performance of users, as well as the maintainability of the code. IDEs use code style information to automatically format the code appropriately, making it easier for programmers to write code that is consistent and adheres to best practices. There are some existing conventions in the world of writing --- every domain has traditions and rules for composing and delivering text, \textit{e.g.}, citation formatting rules, or scientific paper structure. The field could greatly benefit from \WA that are capable of converting plain language text into any required form or convention. Integrating conventions into \WA or even inferring and learning conventions from corpora of texts in a particular organization can help maintain the consistency and readability of texts, especially in the domains of formal and business communication.

\textbf{Explicit context.}
In software development, code is typically placed within a file surrounded by other code it might interact with, the project has a set of used libraries and other parts of the configured environment, and usually there is a history of edits as part of a version control system. All of this is available to a programmer in their IDE. IDEs emphasize this context and suggest code snippets that are relevant to it (for example, suggesting code completions based on the imported libraries), making it easier for programmers to write code quickly that integrates seamlessly with the rest of the project. Bringing context to \WA might help better understand the structure of the text and keep track of changes made to the documents. Using and emphasizing terminology as a context can assist writers in supporting their degree of accuracy and clarity in academic, legal, and technical writing. In turn, version control would be a valuable feature for \WA in any writing domain where collaboration and frequent revisions are expected. While some of this functionality, like version control and document structure visualization, has already been implemented in tools such as Overleaf and Google Docs, other useful applications of these approaches are yet to be developed.  

Applying ideas from IDEs to \WA presents unique challenges that must be conquered. For example, context analysis in IDEs involves algorithms like traversing dependency graphs or analyzing the project structure. In the context of the more ambiguous setting of a general text, it is necessary first to develop a structure-agnostic system using natural language processing techniques to find a relevant context in an array of user files. Additionally, design challenges for implementing new features in \WA are emerging. For instance, how to provide comprehensible multiline suggestion and its various options. While evolving and augmenting a wide range of features and functionality, writing tools would need to have a user-friendly, adaptable interface that allows writers to quickly access and use features without disrupting users' workflow. The tool should not be too distracting or intrusive because that could hinder the writing process and compromise the author's creativity. As a whole, the proposed technology transfer is both rewarding and challenging.

\subsection{IDE as a playground for WA research}

Some modern programming languages such as Python and Swift are designed to be as close to a human language as possible, while also having strict rules and syntax that ensure that the computer can execute the code. Thus, the code could be perceived as an instance of text~\cite{hindle2016naturalness}. Therefore, IDE might be used as a model of \WA to test assumptions about interventions to help people interact with tools efficiently.

A wide variety of data collected and stored in an IDE makes it possible to evaluate and optimize the features of the environment. By measuring, for instance, the readability of the code before and after using some IDE feature, researchers and practitioners can assess the effectiveness of the provided interventions and draw corresponding conclusions. In previous work, this approach has already been tested by using a custom IDE plugin to test hypotheses from the field of human-computer interaction~\cite{davis2022analysis,zyrianov2020automated,kasatskii2023effect,santos2017design,xu2022ide}. The same could be done for topics related to \WA. Although there are certain differences between the mediums, some lessons learned from the development of the IDEs can be useful for the building of the \WA in general and conversely, lessons from WAs can make IDEs more useful.

%% file: sections/03-conclusion.tex
\section{Conclusion}\label{sec:con}

With the development of artificial intelligence, new concerns in the field of \WA are emerging. One of them is that the amount of information processed by the tool and by its users becomes increasingly large and overwhelming, especially for people with neurodiversity. Here we propose that this issue can be addressed by using the decades-long experience of the development of IDE and by using them as a model for future research. 